\documentclass[aps,prd,preprint,tightenlines,superscriptaddress,showpacs,byrevtex,amsmath,amssymb]{revtex4}

\usepackage{graphicx} 
\usepackage{dcolumn}  
\usepackage[dvips]{color}
\usepackage{ulem}

\graphicspath{{ps}}

\newcommand{\ycp}{$y_{CP}$~}
\newcommand{\zdmix}{$D^0$-$\overline{D}{}^0$~}
\newcommand{\zdecay}{$D^0\to K_S^0K^+K^-$~}

\begin{document}

\title{ \quad\\[0.5cm]\boldmath  Measurement of \ycp in $D^0$ meson decays to the $K_S^0K^+K^-$ final state}

\affiliation{Budker Institute of Nuclear Physics, Novosibirsk}
\affiliation{University of Cincinnati, Cincinnati, Ohio 45221}
\affiliation{T. Ko\'{s}ciuszko Cracow University of Technology, Krakow}
\affiliation{Department of Physics, Fu Jen Catholic University, Taipei}
\affiliation{The Graduate University for Advanced Studies, Hayama}
\affiliation{Hanyang University, Seoul}
\affiliation{University of Hawaii, Honolulu, Hawaii 96822}
\affiliation{High Energy Accelerator Research Organization (KEK), Tsukuba}
\affiliation{Hiroshima Institute of Technology, Hiroshima}
\affiliation{Institute of High Energy Physics, Chinese Academy of Sciences, Beijing}
\affiliation{Institute of High Energy Physics, Vienna}
\affiliation{Institute of High Energy Physics, Protvino}
\affiliation{Institute for Theoretical and Experimental Physics, Moscow}
\affiliation{J. Stefan Institute, Ljubljana}
\affiliation{Kanagawa University, Yokohama}
\affiliation{Korea University, Seoul}
\affiliation{Kyungpook National University, Taegu}
\affiliation{\'Ecole Polytechnique F\'ed\'erale de Lausanne (EPFL), Lausanne}
\affiliation{Faculty of Mathematics and Physics, University of Ljubljana, Ljubljana}
\affiliation{University of Maribor, Maribor}
\affiliation{University of Melbourne, School of Physics, Victoria 3010}
\affiliation{Nagoya University, Nagoya}
\affiliation{Nara Women's University, Nara}
\affiliation{National Central University, Chung-li}
\affiliation{National United University, Miao Li}
\affiliation{Department of Physics, National Taiwan University, Taipei}
\affiliation{H. Niewodniczanski Institute of Nuclear Physics, Krakow}
\affiliation{Nippon Dental University, Niigata}
\affiliation{Niigata University, Niigata}
\affiliation{University of Nova Gorica, Nova Gorica}
\affiliation{Novosibirsk State University, Novosibirsk}
\affiliation{Osaka City University, Osaka}
\affiliation{Panjab University, Chandigarh}
\affiliation{Saga University, Saga}
\affiliation{University of Science and Technology of China, Hefei}
\affiliation{Seoul National University, Seoul}
\affiliation{Shinshu University, Nagano}
\affiliation{Sungkyunkwan University, Suwon}
\affiliation{University of Sydney, Sydney, New South Wales}
\affiliation{Tata Institute of Fundamental Research, Mumbai}
\affiliation{Toho University, Funabashi}
\affiliation{Tohoku Gakuin University, Tagajo}
\affiliation{Tohoku University, Sendai}
\affiliation{Department of Physics, University of Tokyo, Tokyo}
\affiliation{Tokyo Metropolitan University, Tokyo}
\affiliation{Tokyo University of Agriculture and Technology, Tokyo}
\affiliation{IPNAS, Virginia Polytechnic Institute and State University, Blacksburg, Virginia 24061}
\affiliation{Yonsei University, Seoul}
  \author{A.~Zupanc}\affiliation{J. Stefan Institute, Ljubljana} 
  \author{I.~Adachi}\affiliation{High Energy Accelerator Research Organization (KEK), Tsukuba} 
  \author{H.~Aihara}\affiliation{Department of Physics, University of Tokyo, Tokyo} 
 \author{K.~Arinstein}\affiliation{Budker Institute of Nuclear Physics, Novosibirsk}\affiliation{Novosibirsk State University, Novosibirsk} 
  \author{V.~Aulchenko}\affiliation{Budker Institute of Nuclear Physics, Novosibirsk}\affiliation{Novosibirsk State University, Novosibirsk} 
 \author{T.~Aushev}\affiliation{\'Ecole Polytechnique F\'ed\'erale de Lausanne (EPFL), Lausanne}\affiliation{Institute for Theoretical and Experimental Physics, Moscow} 
  \author{A.~M.~Bakich}\affiliation{University of Sydney, Sydney, New South Wales} 
  \author{V.~Balagura}\affiliation{Institute for Theoretical and Experimental Physics, Moscow} 
  \author{E.~Barberio}\affiliation{University of Melbourne, School of Physics, Victoria 3010} 
  \author{A.~Bay}\affiliation{\'Ecole Polytechnique F\'ed\'erale de Lausanne (EPFL), Lausanne} 
% ...
\collaboration{The Belle Collaboration}
\noaffiliation
\begin{abstract}
We present a measurement of the \zdmix mixing parameter \ycp using a flavor-untagged 
sample of \zdecay decays. The measurement is based on a 673 fb$^{-1}$ data sample 
recorded with the Belle detector at the KEKB asymmetric-energy $e^+ e^-$ collider. 
Using a method based on measuring the mean decay time for different $K^+K^-$ invariant mass 
intervals, we find $y_{CP} = (+0.11\pm 0.61 ({\rm stat.})\pm 0.52 (\rm syst.))\%$.
\end{abstract}

\pacs{13.25.Ft, 12.15.Ff}

\maketitle

{\renewcommand{\thefootnote}{\fnsymbol{footnote}}}
\setcounter{footnote}{0}

\section{Introduction}

Particle-antiparticle mixing has been observed in the neutral kaon, $B^0_d$ and $B^0_s$ meson systems, and evidence for mixing has recently been found for neutral $D$ mesons.
The mixing occurs through weak interactions 
and gives rise to two distinct mass eigenstates: 
$|D_{1,2}\rangle=p|D^0\rangle\pm~ q|\overline{D}{}^0\rangle$, where $p$ and $q$ are complex coefficients satisfying $|p|^2+|q|^2=1$. The time evolution of the flavor eigenstates, $D^0$ and $\overline{D}{}^0$, is governed by the mixing parameters
$x = (m_1-m_2)/\Gamma$ and $y = (\Gamma_1-\Gamma_2)/2\Gamma$, where $m_{1,2}$ and $\Gamma_{1,2}$ are the masses and widths of the two mass eigenstates $D_{1,2}$, and
$\Gamma = (\Gamma_1+\Gamma_2)/2$.
In the Standard Model (SM) the contribution of the box diagram, successfully describing mixing in the $B$- and $K$-meson systems, is strongly suppressed for $D^0$ mesons due both to the smallness of the $V_{ub}$ element of the Cabibbo-Kobayashi-Maskawa matrix \cite{CKM}, and to the Glashow-Illiopoulos-Maiani mechanism \cite{GIM}. 
The largest SM predictions for the parameters $x$ and $y$, which include the impact of long distance dynamics, are of order $1$\% \cite{mix_th}.
Observation of large mixing could indicate the contribution of new processes and particles.

Evidence for \zdmix mixing has been found in $D^0\to K^+K^-/\pi^+\pi^-$ \cite{Staric,AubertMix}, $D^0\to K^+\pi^-$ \cite{Aubert1,CDF} and $D^0\to K^+\pi^-\pi^0$ \cite{kpipi0} decays. Currently the most precise individual measurements of mixing parameters are those from the relative lifetime difference between $D^0$ decays to $CP$ eigenstates and flavor-specific final states, $y_{CP}$, which equals the parameter $y$ in the limit where $CP$ is conserved. 
Thus far, only $CP$-even final states $K^+K^-$ and $\pi^+\pi^-$ have been used;
the resulting world average value~\cite{hfag} for \ycp is
$(+1.13 \pm 0.27 )\%$. 

In this paper we present a flavor-untagged measurement of $y_{CP}$ using the $CP$-odd component of 
\zdecay decays~\cite{CC}. 
The measurement is performed by comparing mean decay times for different regions of the three-body phase space distribution. As this method does not use a fit to 
the decay time distribution, it does not require detailed knowledge of the 
resolution function or the time distribution of backgrounds. The result has similar statistical sensitivity to that obtained by fitting the decay time distribution. 

\section{Method}
\label{sec_method}
The time-dependent decay amplitude of an initially produced $D^0$ or $\overline{D}{}^0$
can be expressed in terms of the 
neutral $D$ meson amplitudes
$\langle K_S^0K^+K^-|D^0\rangle={\cal A}(s_0,s_+)$ and 
$\langle K_S^0K^+K^-|\overline{D}{}^0\rangle=\overline{\cal A}(s_0,s_+)$,
where $s_0=M_{K^+K^-}^2$ \cite{MassConvention} and $s_{\pm}=M_{K_S^0K^{\pm}}^2$. The explicit expressions are~\cite{LiMing,AsnerCleo}:
\begin{eqnarray}
  \label{eqn1a}
  \langle K_S^0K^+K^-|D^0(t)\rangle&=&
  \frac{1}{2}\!\!\left[{\cal A}(s_0,s_+)+\frac{q}{p}\overline{\cal A}(s_0,s_+)\right]\!\!e_1(t)
  +\frac{1}{2}\!\!\left[{\cal A}(s_0,s_+)-\frac{q}{p}\overline{\cal A}(s_0,s_+)\right]\!\!e_2(t)~~ \\
  \label{eqn1b}
  \langle K_S^0K^+K^-|\overline{D}{}^0(t)\rangle&=&
  \frac{1}{2}\!\!\left[\overline{\cal A}(s_0,s_+)+\frac{p}{q}{\cal A}(s_0,s_+)\right]\!\!e_1(t)
  +\frac{1}{2}\!\!\left[\overline{\cal A}(s_0,s_+)-\frac{p}{q}{\cal A}(s_0,s_+)\right]\!\!e_2(t), ~~
\end{eqnarray}
with $e_{1,2}(t)=\exp\{-i(m_{1,2}-i\Gamma_{1,2}/2)t\}$. In the limit of $CP$ 
conservation ($p/q=1$), Eqs.~(\ref{eqn1a}) and~(\ref{eqn1b}) simplify to:
\begin{equation}
  \langle K_S^0K^+K^-|D^0(t)\rangle= {\cal A}_1(s_0,s_+)e_1(t)+{\cal A}_2(s_0,s_+)e_2(t)
 \label{eqn2a}
\end{equation}
\begin{equation}
  \langle K_S^0K^+K^-|\overline{D}{}^0(t)\rangle= {\cal A}_1(s_0,s_+)e_1(t)-{\cal A}_2(s_0,s_+)e_2(t)\,,
\label{eqn2b}
\end{equation}
where ${\cal A}_1(s_0,s_+)=[{\cal A}(s_0,s_+) +\overline{\cal A}(s_0,s_+)]/2$ and 
${\cal A}_2(s_0,s_+)=[{\cal A}(s_0,s_+) -\overline{\cal A}(s_0,s_+)]/2$. In the isobar model the 
amplitudes ${\cal A}$ and $\overline{\cal A}$ are written as the sum of 
intermediate decay channel amplitudes (subscript $r$) with the same final state, 
${\cal A}(s_0,s_+) = \sum_r a_r e^{i\phi_r}{\cal A}_r(s_0,s_+)$ and
$\overline{\cal A}(s_0,s_+) =  \sum_r \overline{a}_r e^{i\overline{\phi}_r}\overline{\cal A}_r(s_0,s_+) = \sum_r a_r e^{i\phi_r}{\cal A}_r(s_0,s_-)$, where $CP$ conservation in decay has been assumed in the final step.
If $r$ is a $CP$ eigenstate, then ${\cal A}_r(s_0,s_-) = \pm {\cal A}_r(s_0,s_+)$,
where the sign $+$($-$) holds for a $CP$-even(-odd) eigenstate. Hence the amplitude ${\cal A}_1$ is $CP$-even, and the amplitude ${\cal A}_2$ is $CP$-odd.

\begin{figure}[t]
 \includegraphics[width=0.48\textwidth]{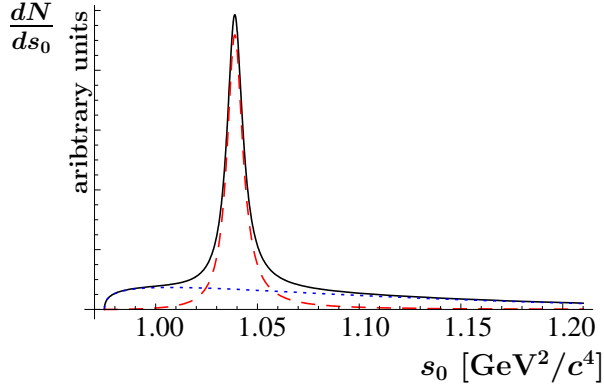}
 \caption{
Projection of time-integrated Dalitz distribution to $s_0$ (solid line), and the $a_1(s_0)$ (dotted line) and $a_2(s_0)$ (dashed line) contributions for the Dalitz model given in \cite{BaBar}.}
 \label{fig_DalitzModel}
\end{figure}
Upon squaring Eqs.~(\ref{eqn2a}) and~(\ref{eqn2b}) we obtain for the time-dependent decay 
rates of initially produced $D^0$ and $\overline{D}{}^{0}$:
\begin{eqnarray}
\frac{dN(s_0,s_+,t)}{dt} & \propto &
  |{\cal A}_1(s_0,s_+)|^2e^{-\frac{t}{\tau}(1+y)} + 
  |{\cal A}_2(s_0,s_+)|^2e^{-\frac{t}{\tau}(1-y)} \nonumber\\
  & & + 2Re[{\cal A}_1(s_0,s_+){\cal A}_2^{\ast}(s_0,s_+)]\cos{\left(x\frac{t}{\tau}\right)}
  e^{-\frac{t}{\tau}} \nonumber\\
  & & + 2Im[{\cal A}_1(s_0,s_+){\cal A}_2^{\ast}(s_0,s_+)]\sin{\left(x\frac{t}{\tau}\right)}
  e^{-\frac{t}{\tau}} \label{eq_mD0}\\
  \frac{d\overline{N}(s_0,s_+,t)}{dt} & \propto &
  |{\cal A}_1(s_0,s_+)|^2e^{-\frac{t}{\tau}(1+y)} + 
  |{\cal A}_2(s_0,s_+)|^2e^{-\frac{t}{\tau}(1-y)} \nonumber\\
  & & - 2Re[{\cal A}_1(s_0,s_+){\cal A}_2^{\ast}(s_0,s_+)]\cos{\left(x\frac{t}{\tau}\right)}
  e^{-\frac{t}{\tau}} \nonumber\\
  & & - 2Im[{\cal A}_1(s_0,s_+){\cal A}_2^{\ast}(s_0,s_+)]\sin{\left(x\frac{t}{\tau}\right)}
  e^{-\frac{t}{\tau}} \label{eq_mD0bar},
\end{eqnarray}
%
where $\tau=1/\Gamma$ is the $D^0$ lifetime. It can be 
shown (see Appendix) that in
the projection of the Dalitz plot onto $s_0$, the last two terms 
in Eqs.~(\ref{eq_mD0}) and~(\ref{eq_mD0bar}) vanish. 
Hence, a projection onto $s_0$  of the time-dependent decay rate
for \zdecay in the limit of $CP$ conservation depends only 
on the mixing parameter $y$:
\begin{eqnarray}
  \frac{dN(s_0,t)}{dt} \propto a_1(s_0) e^{-\frac{t}{\tau}(1+y)} + a_2(s_0) e^{-\frac{t}{\tau}(1-y)}\,,
  \label{dNdt}
\end{eqnarray}
where $a_{1,2}(s_0) = \int{|{\cal A}_{1,2}(s_0,s_+)|^2 ds_+}$. 

Figure~\ref{fig_DalitzModel}
shows the time-integrated projection of the decay rate
(Eq.~\ref{dNdt}) together with the $a_1(s_0)$ and $a_2(s_0)$ contributions;
the plots are obtained using the Dalitz model of Ref.~\cite{BaBar} and 
taking $y=0$. The Dalitz model includes five $CP$-even
intermediate states ($K^0_S a_0(980)^0$, $K^0_S f_0(1370)$, $K^0_S f_2(1270)$, 
$K^0_S a_0(1450)^0$, $K^0_S f_0(980)$), one $CP$-odd intermediate state 
($K^0_S \phi(1020)$) and three flavor-specific intermediate states 
($K^- a_0(980)^+$, $K^- a_0(1450)^+$, $K^+ a_0(980)^-$).

The two terms in Eq.~(\ref{dNdt}) have a different time dependence as well as a different
$s_0$ dependence~(see Fig.~\ref{fig_DalitzModel}). In any given $s_0$ interval, $\cal{R}$, and assuming $y\ll 1$,
the effective $D^0$ lifetime is
\begin{equation}
 \tau_{\cal R}=\tau\left[1+(1-2f_{\cal R})y_{CP}\right],
 \label{effective_tau}
\end{equation}
where $f_{\cal R}=\int_{{\cal R}} a_1(s_0) ds_0/\int_{{\cal R}} (a_1(s_0)+a_2(s_0)) ds_0$, which
represents the effective fraction of the events in the interval ${\cal R}$ due to the ${\cal A}_1$ amplitude. 
In Eq.~(\ref{effective_tau}) we introduced the usual notation $y_{CP}$ for the 
mixing parameter $y$ to indicate that we assumed $CP$ conservation in deriving
Eq.~(\ref{dNdt}). The definition of \ycp in Eq.~(\ref{effective_tau}) is consistent 
with that used in the $D^0\to K^+K^-/\pi^+\pi^-$ measurement \cite{Staric}.

The mixing parameter \ycp can be determined from the relative difference 
in the effective lifetimes of the two $s_0$ intervals, one around the 
$\phi(1020)$ peak (interval ON) and the other in the sideband (interval OFF).
Using Eq.~(\ref{effective_tau}) and taking into account the fact that $[1-(f_{\rm ON}+f_{\rm OFF})]y_{CP}\ll 1$, we obtain
\begin{equation}
  y_{CP}=\frac{1}{f_{\rm ON}-f_{\rm OFF}}\left(\frac{\tau_{\rm OFF}-\tau_{\rm ON}}{\tau_{\rm OFF}+\tau_{\rm ON}}\right).
  \label{eq_ycp}
\end{equation}
The sizes of the ON and OFF intervals are chosen to minimize the statistical
uncertainty on $y_{CP}$. They are determined using the Dalitz model of
\zdecay decays from Ref. \cite{BaBar}. The optimal intervals are found to be:
$M_{K^+K^-} \in [1.015, 1.025]$
GeV/$c^2$ for the ON interval, and 
the union of
 $M_{K^+K^-} \in
[2m_{K^{\pm}},1.010]$ GeV/$c^2$ and $M_{K^+K^-} \in [1.033,1.100]$
GeV/$c^2$ for the OFF interval.

\section{Measurement}

This section is organized as follows: in subsection~\ref{sec_A} we
         describe how signal decays are reconstructed; in 
         subsection~\ref{sec_B} we describe how the mean decay time 
         of the signal is extracted in the presence of background; 
        in subsections~\ref{sec_C} and \ref{sec_D} we describe how the background 
        fraction and mean lifetime, respectively, are determined;
        in subsection~\ref{sec_E} we describe how $f_{\rm ON}-f_{\rm OFF}$ is determined;
        and in subsection~\ref{sec_F} we give the result for $y_{CP}$.

\subsection{Reconstruction of events}\label{sec_A}
The data were recorded with the Belle detector at the KEKB asymmetric-energy $e^+e^-$ collider \cite{KEKB}. 
The Belle detector consists of a silicon vertex detector (SVD), a 50-layer central drift chamber (CDC), an array of
aerogel threshold Cherenkov counters (ACC), a barrel-like arrangement of time-of-flight scintillation counters (TOF), and an electromagnetic calorimeter
(ECL) comprised of CsI(Tl) crystals located inside a superconducting solenoid coil that provides a 1.5~T magnetic field.  An iron flux-return located outside of
the coil is instrumented to detect $K_L^0$ mesons and to identify muons (KLM). The detector is described in detail elsewhere~\cite{Belle}.
Two inner detector configurations were used. A 2.0 cm beampipe and a 3-layer silicon vertex detector was used for the first sample
of 156 fb$^{-1}$, while a 1.5 cm beampipe, a 4-layer silicon detector and a small-cell inner drift chamber were used to record  
the remaining 517 fb$^{-1}$ of data \cite{svd2}. We use an EvtGen- \cite{evtgen} and GEANT-based \cite{GEANT} Monte Carlo (MC) simulated sample, in which the number of reconstructed events is about three times larger than in the data sample, to study the detector response.

The $K_S^0$ candidates are reconstructed in the $\pi^+\pi^-$ final state. We require that 
the pion candidates form a common vertex with a $\chi^2$ fit probability of at least $10^{-3}$, 
and that they be displaced from the $e^+e^-$ interaction point (IP) by at least 0.9~mm 
in the plane perpendicular to the beam axis. 
We also require that they
have an invariant mass $M_{\pi^+\pi^-}$ in the interval [$0.468$, $0.526$]~GeV/$c^2$. 
We reconstruct $D^0$ candidates by combining the $K_S^0$ candidate with two oppositely charged tracks assumed to be kaons. We require charged kaon candidate tracks to satisfy particle identification criteria based upon $dE/dx$ ionization energy loss
in the CDC,
time-of-flight, and Cherenkov light yield in the ACC \cite{BellePID}. These tracks are required to have at least one SVD hit in both $r-\phi$ and $z$ coordinates. A $D^0$ momentum greater than 2.55~GeV/$c$ in the $e^+e^-$ CM frame is required to reject $D$ mesons produced in $B$-meson decays and to suppress combinatorial background. Events with a
$K_S^0K^+K^-$ invariant mass ($M_{K_S^0K^+K^-}$) in the 
interval [$1.77$, $1.96$]~GeV/$c^2$ are retained for further analysis.. 

The proper decay time of the $D^0$ candidate is calculated 
by projecting the vector joining the production and decay vertices, $\vec{L}$, onto 
the $D^0$ momentum vector $\vec{p}_D$: 
$t=(m_{D^0}/p_D)\vec{L}\cdot(\vec{p}_D/p_D)$, where $m_{D^0}$ is the nominal $D^0$ mass. 
Charged and neutral kaon candidates are required to originate from a common vertex for which the $\chi^2$ fit probability is larger than $10^{-3}$. 
According to 
simulation studies,
if the $D^0$ decay position is determined by fitting
the two prompt charged tracks to a common vertex,
the decay length and the opening angle of the $K^+$ and $K^-$
(and thus their invariant mass) are strongly correlated.
This correlation is avoided by determining the $D^0$ decay length from a fit where only a single charged kaon and the $K^0_S$ are fitted to a common vertex. 
Both $K^{\pm}K^0_S$ vertex combinations are required to have a
$\chi^2$ probability larger than $10^{-3}$; for the $\vec{L}$ determination, 
the one with the higher $\chi^2$ fit probability is chosen.
The $D^0$ production point is taken to be the intersection of the trajectory of the $D^0$ candidate with the IP region. The average position of the IP is calculated for every ten thousand events from the primary vertex distribution of hadronic events. The size of the IP region is typically $3.5$ mm in the direction of the beam, 100 $\mu$m in the horizontal direction,
and 5 $\mu$m in the vertical direction. 
The uncertainty in a $D^0$'s candidate's proper decay time ($\sigma_t$) is evaluated from 
the corresponding covariance matrices. We require $\sigma_t<600$ fs.  
The maximum of the $\sigma_t$ distribution is at $\sim230$~fs.

\begin{figure}[t!]
 \begin{center}
 \includegraphics[width=0.8\textwidth]{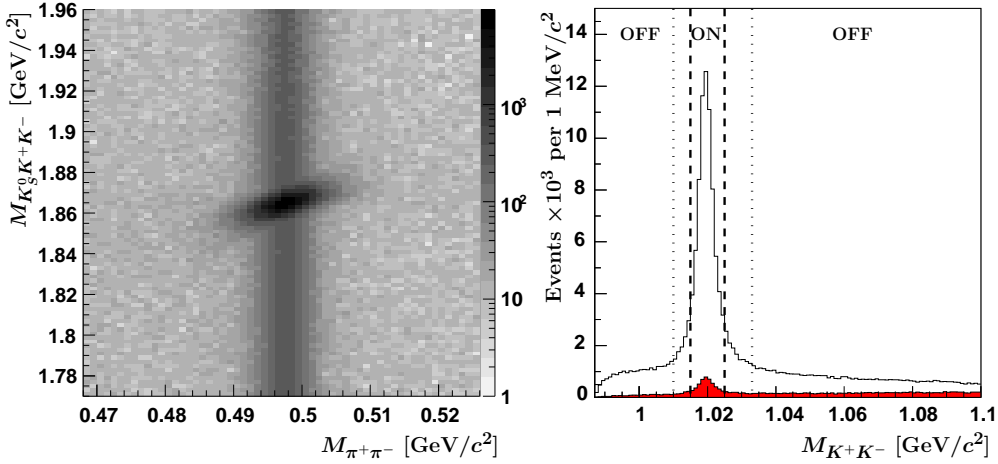}
 \end{center}
 \caption{(Left) $(M_{\pi^+\pi^-},M_{K^0_SK^+K^-})$ distribution of selected events. (Right) $M_{K^+K^-}$ distribution of events in the $|M_{K^0_SK^+K^-}-m_{D^0}|<10$~MeV/$c^2$ and $|M_{\pi^+\pi^-}-m_{K^0_S}|<10$~MeV/$c^2$ 
region (unfilled histogram), and $20<|M_{K^0_SK^+K^-}-m_{D^0}|<30$~MeV/$c^2$ and 
$|M_{\pi^+\pi^-}-m_{K^0_S}|<10$~MeV/$c^2$ region (filled histogram). 
Dashed (Bottom) vertical lines indicate the boundaries of ON (OFF) intervals.}
 \label{fig_md0mksmkk}
\end{figure}

Around $362\times 10^{3}$ events pass all selection criteria. The $(M_{\pi^+\pi^-},M_{K_S^0K^+K^-})$ and $M_{K^+K^-}$ distributions of these events 
are shown in Fig.~(\ref{fig_md0mksmkk}). 

\subsection{Effective signal lifetime}\label{sec_B}

We determine the effective lifetime of \zdecay decays from the 
distribution of proper decay times as follows.
The proper decay time distribution of $D^0$ candidates can be parameterized as:
\begin{equation}
 {\cal P}(t)=p\frac{1}{\tau}\int e^{-t'/\tau}\cdot R(t-t',t_0)dt'+(1-p)B(t),
 \label{eq_pdt}
\end{equation}
where the first term represents the measured distribution of signal events with lifetime $\tau$, convolved with a resolution function, $R(t,t_0)$; $t_0$ corresponds to a possible shift of the resolution function from zero, $p=N_s/(N_s+N_b)$ is the fraction of signal events, and the last term, $B(t)$, describes the distribution of background events. 
Since the average of the convolution is the sum of the averages of the convolved functions, we can express the lifetime of signal events in region ${\cal R}$ (shifted for the resolution function offset) as 
\begin{equation}
\tau_{\cal R} + t_0^{\cal R} = \frac{\langle t\rangle^{\cal R}-(1-p^{\cal R})\langle t\rangle^{\cal R}_b}{p^{\cal R}},
 \label{eq_taus}
\end{equation}
where $\langle t\rangle^{\cal R}$ and $\langle t\rangle^{\cal R}_{b}$ are the mean proper decay times of all events and background events, respectively. 
By measuring $\langle t\rangle^{\cal R}$ and $\langle t\rangle^{\cal R}_{b}$ for events in ON and OFF intervals of $M_{K^+K^-}$ we can obtain the two effective lifetimes and $y_{CP}$ from Eq.~(\ref{eq_ycp}). 
Note that the resolution function offset, $t_0$, if small ($t_0\ll\tau$) and equal in ON and OFF regions, introduces a negligible bias ($\approx y_{CP}\cdot t_0/\tau$) in the measurement, since it cancels in the numerator of Eq.~(\ref{eq_ycp}). We use the simulated sample to confirm
that the resolution function offsets $t_0^{\rm ON}$ and $t_0^{\rm OFF}$ are equal to within the statistical uncertainty. 

The requirement of minimal $K_S^0$ candidate flight distance introduces a bias in the reconstructed mean proper decay time of signal \zdecay decays: events where both $D^0$ and $K_S^0$ candidates are short-lived are rejected by this requirement. This introduces an $+0.5\%$ bias in the mean of the measured proper decay times for $D^0\to K_S^0K^+K^-$; the effect on the $y_{CP}$ parameter is smaller and is included in the systematic error.

\subsection{Signal and background fractions}\label{sec_C}

Signal and background fractions are determined from a fit to the distribution 
of events in the $(M_{\pi^+\pi^-},M_{K^0_SK^+K^-})$ plane. 
In order to model the correlation between invariant masses $M_{K^0_SK^+K^-}$ 
and $M_{\pi^+\pi^-}$ of signal events (see Fig.~\ref{fig_md0mksmkk}),
we parameterize the signal shape by a rotated triple two-dimensional
Gaussian distribution. The individual Gaussians 
are required to have the same mean value,
which is allowed to vary in the fit. The ratio of the Gaussian widths is 
fixed to the Monte Carlo (MC) simulated value,
and only the width of the core Gaussian and the three 
correlation coefficients are left free. 
\begin{figure}[t!]
 \begin{center}
 \includegraphics[width=1\textwidth]{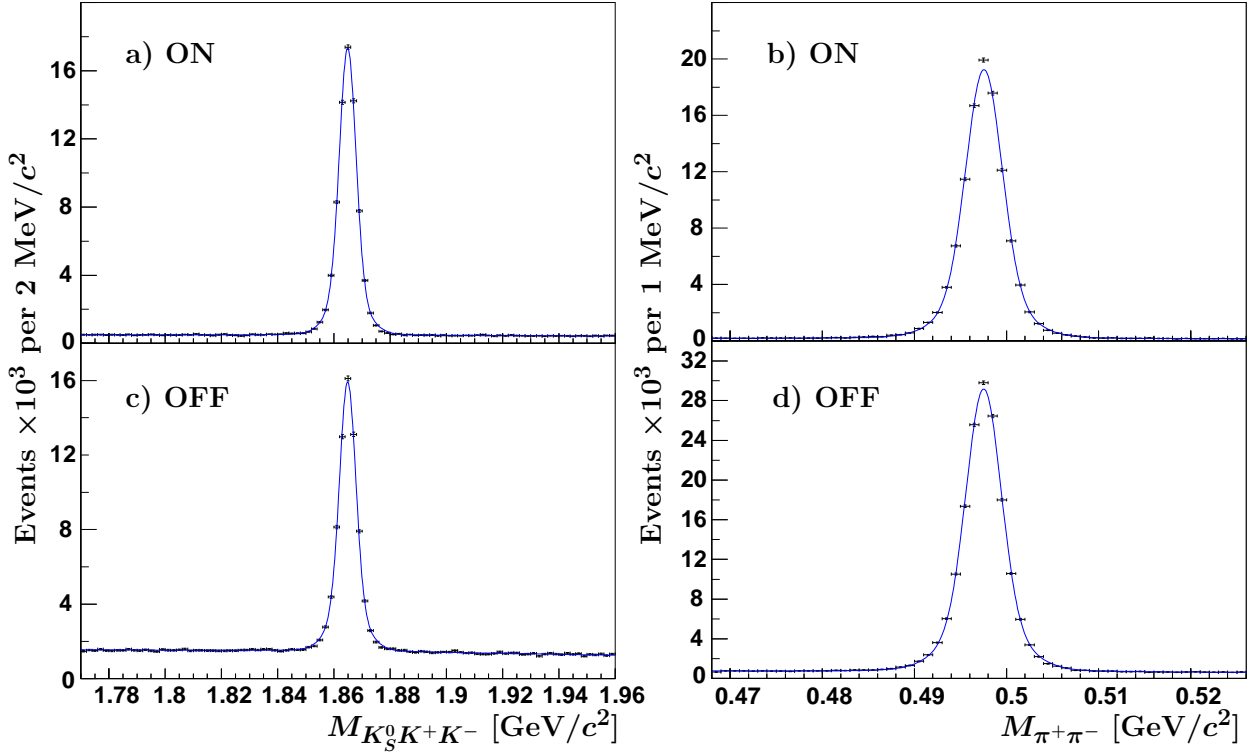}
 \end{center}
 \caption{Invariant masses $M_{K^0_SK^+K^-}$ (a, c) and $M_{\pi^+\pi^-}$ (b, d) of events passing all selection criteria for ON and OFF intervals in $M_{K^+K^-}$. Superimposed on the data (points with error bars) are results of the fit (solid line).}
 \label{fig_md0mksfit}
\end{figure}

Background events are classified into three categories according to their 
distribution in the $(M_{\pi^+\pi^-},M_{K^0_SK^+K^-})$ plane (see Fig.~\ref{fig_md0mksmkk}): 
true $K_S^0$ background, $D^0 \to K^+K^-\pi^+\pi^-$ decays with the pion pair not 
originating from a $K_S^0$, and remaining background. True $K_S^0$ background events are random combinations of charged kaons with correctly reconstructed $K^0_S$ candidates; the shape in $M_{\pi^+\pi^-}$ is fixed to be the same as signal while in $M_{K^0_SK^+K^-}$ it is parameterized with a second-degree polynomial. 
The remaining background events are random combinations of charged particles and are parameterized as a polynomial of first degree in $M_{\pi^+\pi^-}$ and second degree in $M_{K^0_SK^+K^-}$. The $D^0\to K^+K^-\pi^+\pi^-$ decays are peaking in $M_{K^0_SK^+K^-}$, but not in $M_{\pi^+\pi^-}$. According to MC simulation,
the contribution of these events is small ($\sim 0.1\%$);
thus they are not included in the fit but considered as a systematic uncertainty.

The fractions and shapes are determined in a three-step fit for both 
ON and OFF regions. First, the fraction of signal events $F_{\rm sig}$
is obtained from a fit to the one-dimensional projection in $M_{K^0_SK^+K^-}$. 
In the second step, we fit the projection in $M_{\pi^+\pi^-}$ to find the sum of the fractions 
of signal and true $K^0_S$ events, $F_{\rm sig}+F_{\rm tKS}$. 
Finally, we determine 
the signal shape parameters from a two-dimensional fit in which 
we use the $F_{\rm sig}$ and $F_{\rm tKS}$ results from the previous steps. 
The fitting procedure was checked using a high-statistics sample of 
simulated signal and background events and found to correctly 
reproduce the true event fractions.

\begin{figure}[t!]
 \includegraphics[width=0.5\textwidth]{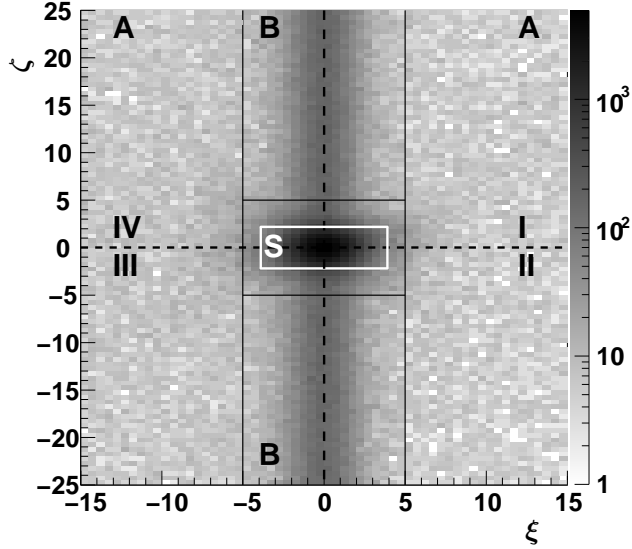}
 \caption{The $(\xi,\zeta)$ distribution of selected events. Signal box S, and sideband regions A and B are defined in the text. Quadrants denoted by I - IV are used in the systematic uncertainty estimate, as described in section~\ref{sec_syst}.}
\label{fig_sidebands}
\end{figure}

The results of this procedure
are shown in Fig.~\ref{fig_md0mksfit}. We find $(72.3\pm0.4)\times 10^3$ signal events
in the ON region and $(62.3\pm 0.7)\times 10^3$ events in the OFF region. To achieve the best statistical accuracy
on the $y_{CP}$ measurement, we optimize the size of the signal box. Because the
invariant masses $M_{K^0_SK^+K^-}$ and $M_{\pi^+\pi^-}$ are correlated for signal events, we define the
signal box in the rotated variables:
\begin{eqnarray}
 \xi & = & \frac{M_{\pi^+\pi^-}-M_{K^0_S}}{\sigma_{K^0_S}}\\
 \zeta & = & \frac{\rho}{\sqrt{1-\rho^2}}\xi-\frac{1}{\sqrt{1-\rho^2}}\frac{M_{K^0_SK^+K^-}-M_{D^0}}{\sigma_{D^0}},
\end{eqnarray}
where $M_{K^0_S}=497.533\pm 0.005$~MeV$/c^2$ and $M_{D^0}=1864.874\pm 0.009$~MeV$/c^2$ are fitted $K^0_S$ and $D^0$ masses, $\sigma_{K^0_S}=1.880\pm 0.008$~MeV$/c^2$ and $\sigma_{D^0}=2.839\pm0.014$~MeV$/c^2$ are widths of the core Gaussian function, 
and $\rho=0.571\pm 0.003$ is the correlation coefficient. The uncertainties are 
statistical only. The signal region that minimizes the statistical uncertainty on 
$y_{CP}$ (signal box) is found to be 
$|\xi|<3.9$ and $|\zeta|<2.2$. The two-dimensional distribution of $(\xi,\zeta)$ for the selected data is shown in Fig.~\ref{fig_sidebands}. The signal fractions in the signal box are $(96.94\pm0.06)$\% and $(90.53\pm0.16)$\% in the ON and OFF intervals, respectively.

The fraction of $D^0 \to K^+K^-\pi^+\pi^-$ decays in the signal box is estimated
by fitting the $M_{K^0_SK^+K^-}$ projection for events in the sideband regions $M_{\pi^+\pi^-} < 0.480$~GeV/$c^2$ and $M_{\pi^+\pi^-} > 0.514$~GeV/$c^2$, where the contributions of signal and true $K_S^0$ background are small. 
The fractions of this background extrapolated to the signal box are found to be $(0.02\pm0.01)\%$ and $(0.07\pm0.02)\%$ in the ON and OFF intervals, respectively, and
are reproduced well by MC simulation.

\subsection{Mean proper decay time of background events}\label{sec_D}
The mean proper decay time of background inside the signal box, $\langle t\rangle_b$, is determined
from sideband regions A and B in the $(\xi,\zeta)$ plane as shown in Fig.~\ref{fig_sidebands}. 
The regions are chosen larger than the signal box 
to minimize the uncertainty on
$\langle t\rangle_b$.
To an excellent approximation, the  
mean proper decay times in sideband regions A and B ($\langle t\rangle^{\rm A}$ and 
$\langle t\rangle^{\rm B}$) can be expressed as
\begin{eqnarray}
\langle t\rangle^{\rm A}  & = & p_{\rm tKs}^{\rm A}\langle t \rangle_{\rm tKs} + p_{\rm rest}^{\rm A}\langle t \rangle_{\rm rest}, \label{eq_taubA}\\
\langle t\rangle^{\rm B}  & = & p_{\rm tKs}^{\rm B}\langle t \rangle_{\rm tKs} + p_{\rm rest}^{\rm B}\langle t \rangle_{\rm rest}, \label{eq_taubB}
\end{eqnarray}
where $p_{\rm tKs}^{\rm A(B)}$ and $p_{\rm rest}^{\rm A(B)}$ are the fractions of true $K_S^0$ and 
the remaining background in region A (B). Similarly, the mean proper decay time of
background in the signal box S can be expressed as
\begin{eqnarray}
  \langle t\rangle_b = p_{\rm tKs}^{\rm S} \langle t\rangle_{\rm tKs} + p_{\rm rest}^{\rm S} \langle t\rangle_{\rm rest}
  \label{eq_taubS}
\end{eqnarray}

By solving Eqs.~(\ref{eq_taubA}) and~(\ref{eq_taubB}) for $\langle t\rangle_{\rm tKs}$ and $\langle t\rangle_{\rm rest}$, and inserting
the results into Eq.~(\ref{eq_taubS}), we obtain
\begin{equation}
\langle t \rangle_{\rm b}=\frac{P^{\rm S}\left(\langle t \rangle^{\rm A}-\langle t \rangle^{\rm B}\right)+P^{\rm A}\langle t \rangle^{\rm B}-P^{\rm B}\langle t \rangle^{\rm A}}{P^{\rm A}-P^{\rm B}},
\label{eq_taubkg_sigbox}
\end{equation}
where $P^i = p_{\rm tKs}^i/(p_{\rm tKs}^i+p_{\rm rest}^i),~i={\rm A,B,S}$. The fractions $p_{\rm tKs}^i$ and 
$p_{\rm rest}^i$, $i={\rm A,B,S}$ are calculated from the results of the two-dimensional fit
discussed in the previous section. In Table~\ref{tab_tbkg} we 
list the
quantities used in Eq.~(\ref{eq_taubkg_sigbox}) and the resulting 
$\langle t \rangle_b$ for regions ON and OFF.

In deriving Eq.~(\ref{eq_taubkg_sigbox}), we have assumed that in regions A, B, and S the mean proper
decay times $\langle t \rangle_{\rm tKs}$ and $\langle t \rangle_{\rm rest}$ are equal. This assumption has been validated
using MC simulation. We have also neglected the signal leakage into regions A and B; if we compare, using MC simulation, the mean proper decay time of background events
found in the signal box with that calculated from Eq.~(\ref{eq_taubkg_sigbox}); we find
agreement well within one standard deviation. 
The small deviations due to these assumptions are
included in the systematic uncertainty.
\begin{table*}[t!]
\begin{center}
\caption{Mean proper decay times of events populating sideband regions A and B in the $(\xi,\zeta)$ plane, $\langle t \rangle^{\rm A}$ and $\langle t \rangle^{\rm B}$, fractions $P^i$ ($i=$ S, A, B) and estimated mean proper decay times of background events, $\langle t \rangle_{b}$, populating the signal box, for events in the ON and OFF intervals in $M_{K^+K^-}$. The uncertainties are statistical only.}
\label{tab_tbkg}
\begin{tabular}{c|cc||ccc||c}\hline\hline
$M_{K^+K^-}$ & $\langle t \rangle^{\rm A}$ [fs] & $\langle t \rangle^{\rm B}$ [fs] & $P^{\rm S}$ [\%]     & $P^{\rm A}$ [\%]     & $P^{\rm B}$ [\%]     & $\langle t \rangle_{b}$ [fs] \\\hline
ON           & $~223\pm14$ & $\phantom{1}63.6\pm4.7$ & $93.31\pm0.41$ & $7.2\pm1.8$    & $91.83\pm0.23$ & $\phantom{1}60.8\pm4.8$     \\ 
OFF          & $237.7\pm7.4$ & $140.3\pm3.1$ & $90.17\pm0.32$ & $5.1\pm1.2$    & $88.02\pm0.17$ & $137.8\pm3.2$    \\\hline
\end{tabular}
\end{center}
\end{table*}

\subsection{Fit to the \boldmath$s_0$ distribution}\label{sec_E}
The ${\cal A}_1$ fractions, $f_{ON}$ and $f_{OFF}$, are obtained from a fit to the $s_0$ distribution. We use two different
Dalitz models of \zdecay decays to parameterize the distribution: 
a four-resonance model from Ref.~\cite{BaBarII}, and an eight-resonance model 
from Ref.~\cite{BaBar}.
The main sensitivity to $y_{CP}$ arises from $K_S^0\phi(1020)$ and $K_S^0a_0(980)^0$ 
intermediate states, since the two have opposite $CP$ eigenvalues. Because all 
resonance parameters cannot be determined from a one-dimensional fit,
we fix the parameters of the resonances with smaller fit fractions using the amplitudes and phases 
from the corresponding model and world averages for masses and widths; we vary 
only the amplitudes of $K_S^0\phi(1020)$ and $K^-a_0(980)^+$ (four-resonance model) 
or the amplitudes of $K_S^0\phi(1020)$ and
$K^-a_0(1450)^+$ (eight-resonance model), mass and width of the $\phi(1020)$, 
and the coupling constant $g_{KK}$ of the Flatte parameterization of the $a_0(980)^0$.

The signal distribution is parameterized as
\begin{equation}
 {\cal F}_{\rm s}(s_0)=\varepsilon(s_0)\int \varepsilon(s_+)|{\cal A}_1(s_0,s_+)+{\cal A}_2(s_0,s_+)|^2ds_+,
\end{equation}
where $\varepsilon$ is the reconstruction efficiency determined from a sample of MC 
events in which the decay mode was generated according to phase space; the efficiency is
found to be factorizable in the Dalitz variables $s_0$ and $s_+$. 
The background parameterization is obtained from the sideband region $5<|\zeta|<25$, where $|\xi|<3.9$ corresponds to the signal region.
A $\chi^2$ test of the MC $s_0$ distributions of background events from the signal and sideband regions yields $\chi^2=88.9$ for $99$ degrees of freedom;
thus we conclude that the $s_0$ distribution of events taken from the sideband region
satisfactorily describes the background distribution in the signal box.

\begin{figure}[t!]
 \begin{center}
 \includegraphics[width=0.8\textwidth]{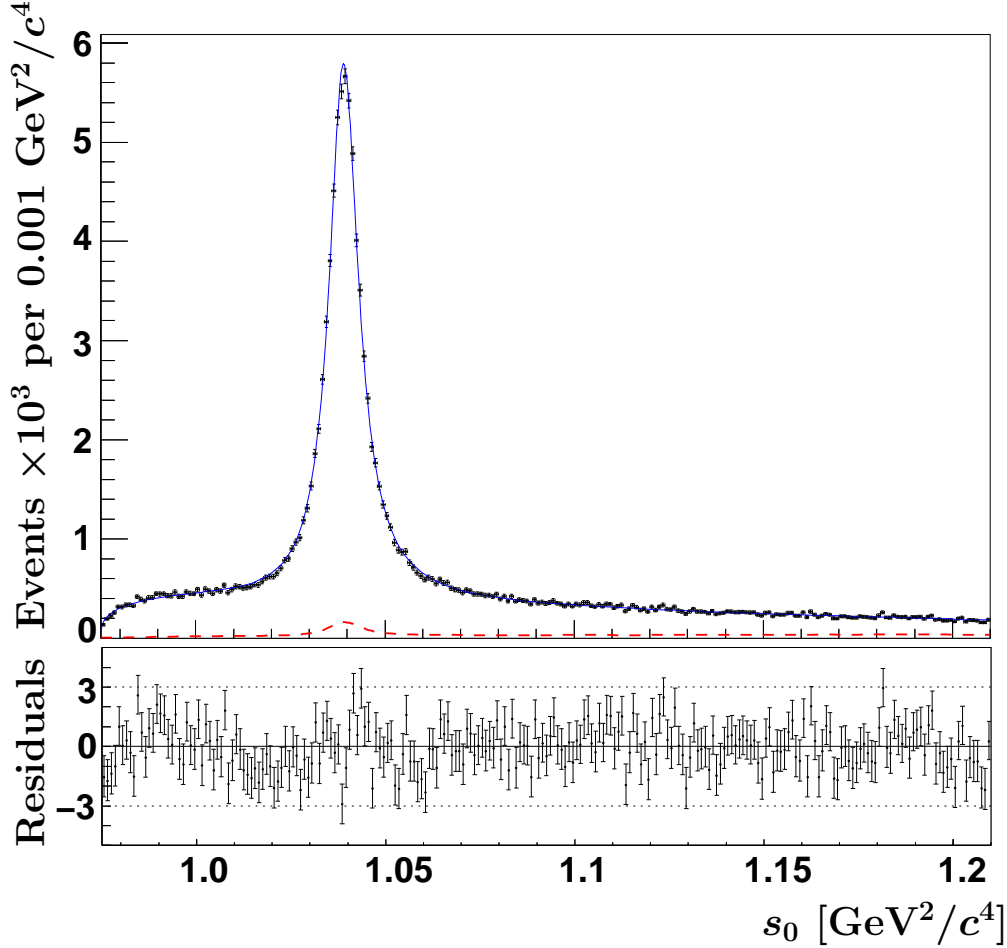}
 \end{center}
 \caption{The $s_0$ distribution of \zdecay decays with superimposed fit results for the 8-resonance Dalitz model given in Ref. \cite{BaBar}. The solid curve is the overall fitted function and the dashed curve represents the background contribution. }
 \label{fig_s0_untaggedfits}
\end{figure}

Figure~\ref{fig_s0_untaggedfits} shows fit results for the eight-resonance model, which we use to determine the fraction difference $f_{\rm ON}-f_{\rm OFF}$, since it provides a better description of the $s_0$ distribution. The reduced $\chi^2$ is 1.28 for the eight-resonance model and 1.91 for the four-resonance model for 230 degrees of freedom.
In Table~\ref{tab_fractionsBest} the fraction differences $f_{\rm ON}-f_{\rm OFF}$ are 
given for both Dalitz models.
The left column lists the values calculated from the data in Refs.~\cite{BaBarII,BaBar},
and the right column lists the values calculated from the results of our fit.
Uncertainties in $f_{\rm ON}-f_{\rm OFF}$ are calculated using the statistical 
errors of amplitudes and phases, without taking into account any correlation 
between them.
Although the models are different, with distinct resonant structure \cite{model_diff}, 
the differences $f_{\rm ON} - f_{\rm OFF}$ calculated for the two models are very similar. 
The small difference between them is included as a systematic uncertainty.

\begin{table}[t!]
\begin{center}
\caption{Fraction difference $f_{\rm ON} - f_{\rm OFF}$ for the two Dalitz models. The nominal values are
calculated from the data in Refs.~\cite{BaBarII,BaBar}, and the fitted values from our fit results.}
\label{tab_fractionsBest}
 \begin{tabular}{c|c|c}\hline \hline
                         & \multicolumn{2}{c}{$f_{\rm ON} - f_{\rm OFF}$}\\ \hline
 Model                   & Nominal  & Fitted\\
 \hline
 4 res. \cite{BaBarII}   & $-0.730\pm0.031$       & $-0.732\pm0.002$\\ 
 8 res. \cite{BaBar}     & $-0.753\pm0.004$       & $-0.769\pm0.005$\\
 \hline
\end{tabular}
\end{center}
\end{table}

\subsection{Results}\label{sec_F}
\begin{figure}[t!]
 \begin{center}
 \includegraphics[width=0.4\textwidth]{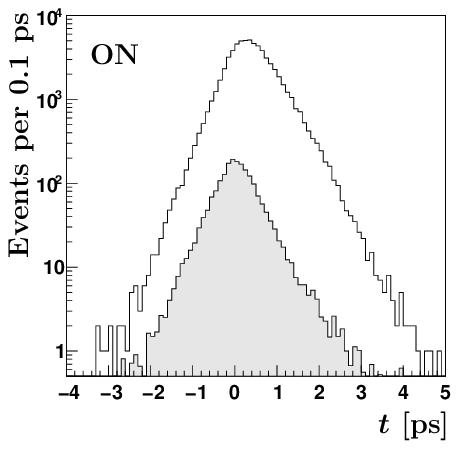}
 \includegraphics[width=0.4\textwidth]{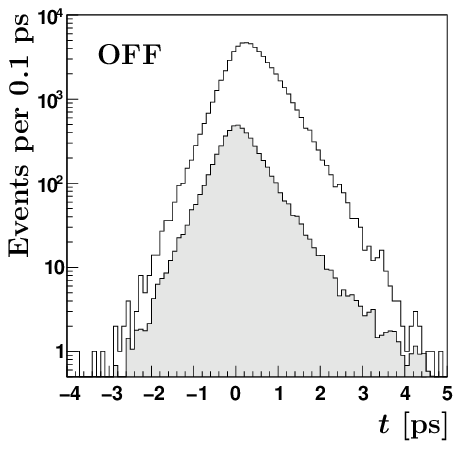}
 \end{center}
 \caption{The proper decay time distributions of all events (unfilled histogram) and 
background events (hatched histogram) populating the signal box S for the ON and OFF intervals.}
 \label{fig_dataPDT}
\end{figure}
\begin{table}[t!]
\begin{center}
\caption{Measured mean proper decay times in the signal box $\langle t\rangle$, effective background lifetimes $\langle t\rangle_b$, signal fractions $p$ and the resulting effective signal lifetimes. The uncertainties are statistical only.}
\label{tab_untagged_DATA}
\begin{tabular}{c|cc|c||c}\hline\hline
$M_{K^+K^-}$ & $\langle t \rangle$ [fs]  & $\langle t \rangle_{b}$ [fs] & $p$ [\%]    &  $\tau+t_0$ [fs]\\ \hline
ON           & $402.7\pm2.5$ & $60.8\pm4.8$  & $96.94\pm 0.06$    & $413.4\pm 2.5$       \\ 
OFF          & $386.7\pm2.6$ & $137.8\pm3.2$ & $90.53\pm 0.16$    & $412.7\pm 3.0$       \\\hline
\end{tabular}
\end{center}
\end{table}

Figure~\ref{fig_dataPDT} shows the proper decay time distributions of 
selected events in the signal box S for the ON and OFF intervals. The 
distribution of background events is estimated from proper decay time 
distributions of events populating the sideband regions A and B and 
the known fractions of the true $K^0_S$ background and the remaining 
background in all three regions.
Inserting the values for $\langle t\rangle$, $\langle t\rangle_b$, and $p$
(the fraction of signal) 
into Eq.~(\ref{eq_taus}) yields 
$\tau_{ON}+t_0^{ON}=(413.4\pm 2.5)$~fs and
$\tau_{OFF}+t_0^{OFF}=(412.7\pm 3.0)$~fs. 
These results are summarized in Table~\ref{tab_untagged_DATA}.
The measured values for $\tau+t_0$ are close to the world average for $\tau_{D^0}$, and, since $y_{CP}\ll 1$, this implies $t_0/\tau$ is $\sim 1\%$ or less. Since the topology of events in the ON and OFF intervals is almost identical, 
we assume $t_0^{ON}=t_0^{OFF}$ and include a systematic error to account
for possible deviations from this assumption. This leads to a normalized lifetime difference 
$(\tau_{\rm OFF}-\tau_{\rm ON})/(\tau_{\rm OFF}+\tau_{\rm ON})=(-0.09\pm 0.47)\%$ between the two regions, where the uncertainty is statistical only. 
The difference in the ${\cal A}_1$ fraction corresponding to the eight-resonance model (see Table~\ref{tab_fractionsBest}) is
$f_{ON}- f_{OFF}=(-0.769\pm0.005)$; 
therefore, from Eq.~(\ref{eq_ycp}) we obtain $y_{CP}=(+0.11\pm0.61(\rm stat.))\%$.

\section{Systematics}\label{sec_syst}

We consider separately systematic uncertainties arising from experimental 
sources and from the \zdecay decay model. 
First, we check the simulated sample to confirm
that the resolution function offsets $t_0^{\rm ON}$ and $t_0^{\rm OFF}$ are equal. 
The small difference observed is consistent with the statistical error but 
conservatively propagated to \ycp and taken as a systematic uncertainty ($\pm0.38$\%).

The mean proper decay time 
of background events populating the 
signal box (calculated from Eq.~(\ref{eq_taubkg_sigbox})) assumes a 
negligible contribution of signal events in sideband regions A and B, and also
assumes equal mean proper decay times of the two background categories in all three 
regions, A, B and S. The systematic uncertainty resulting from the first
assumption is evaluated by including the small residual fraction of signal 
events in regions A and B in the $\langle t \rangle_{b}$ calculation; the
resulting change in $y_{CP}$ is $\pm0.01\%$. The uncertainty resulting from the second assumption is evaluated by MC simulation; mean proper decay times of the two background categories are found to be consistent within statistical uncertainty in all three regions. Small differences between the mean proper decay times of the two background categories in the S, A and B regions result in $\pm0.09\%$ and $0.04\%$ variations of $y_{CP}$ for true $K_S^0$ and remaining background, respectively. We add in quadrature the above three contributions to obtain a $\pm 0.10\%$ systematic error on $y_{CP}$.

The contribution of $D^0\to K^+K^-\pi^+\pi^-$ decays in our sample is found to be small 
 and thus is not included. We evaluate their effect on \ycp by taking the 
 fraction of these events in the ON and OFF intervals from data, and their mean proper decay time from the simulated sample. The resulting change
 in \ycp is $\pm0.07\%$. We include this change in the systematic uncertainty.

We study the choice of sideband regions used to determine $\langle t \rangle_b$
as follows. The sidebands A and B are divided into four subregions (denoted I-IV) as
shown in Fig.~\ref{fig_sidebands}. The mean proper decay time of background 
events is then calculated using events in subregions (I,III) or (II,IV), and a difference 
of 0.05\% in \ycp is observed. This change is included as a systematic uncertainty.

Possible systematic effects of selection criteria are studied by varying 
the signal box size and the selection criteria for $\sigma_t$ and the $K^0_S$ 
flight distance. Although no statistically significant deviation is observed, 
the maximum difference in \ycp is (conservatively) assigned as a 
systematic uncertainty ($\pm0.30$\%). 

The fitting procedure is tested using the simulated sample. A small 
difference between the fitted and 
true fractions of signal events in the signal box is propagated to \ycp 
and included as a systematic uncertainty ($\pm0.10$\%).

The mean proper decay times of events populating the signal 
box S and the sideband regions A and B are taken to be the means 
of histograms of the proper decay times for events 
populating these regions. Changing the binning and intervals used 
in these histograms over a wide range results in a change in \ycp of
$\pm0.07\%$; we include this as an additional systematic uncertainty.

Finally, we estimate the systematic uncertainty due to our 
choice of \zdecay decay model. First, we compare the fraction 
difference $f_{\rm ON} - f_{\rm OFF}$ obtained using the four- and eight-resonance 
Dalitz models. Despite the difference between the models in their
resonant substructure~\cite{model_diff}, the values for $f_{\rm ON} - f_{\rm OFF}$ 
are similar (see Table~\ref{tab_fractionsBest}). We assign a 3\% relative error 
to $y_{CP}$ due to the small difference in the above fractions. An additional 2\% 
relative error is assigned due to the small difference between the fitted and 
nominal values of $f_{\rm ON} - f_{\rm OFF}$.
If the reconstruction efficiency $\varepsilon(s_+)$ were constant, the contribution of the real and 
imaginary parts of the interference term ${\cal A}_1{\cal A}_2^{\ast}$ in 
Eq.~(\ref{eq_mD0}) would vanish after 
integrating over $s_+$. A slight decrease of $\varepsilon(s_+)$ near 
the kinematic boundaries is observed from a large sample of simulated 
events; the effect of this variation on \ycp is studied and found to be negligible.

Adding all decay-model systematic uncertainties in quadrature 
with the statistical uncertainty in $f_{ON}- f_{OFF}$ ($=-0.769\pm0.005$,
see Table~\ref{tab_fractionsBest}) yields a total uncertainty 
due to the decay model of 0.01\%. Combining this in quadrature with all other sources of systematic 
uncertainty gives a total systematic error on \ycp of 0.52\%. The 
individual contributions to the total systematic error are listed in 
Table~\ref{tab_syst}.
\begin{table*}[t!]
\caption{Sources of the systematic uncertainty on $y_{CP}$.}
\label{tab_syst}
\begin{tabular}
{@{\hspace{0.5cm}}l@{\hspace{0.5cm}}|@{\hspace{0.5cm}}c@{\hspace{0.5cm}}}
\hline \hline
Source & Systematic error (\%) \\
\hline
Resolution function offset difference $t^{\rm OFF}_0-t^{\rm ON}_0$ & $\pm 0.38$ \\
Estimation of $\langle t \rangle_{b}$                              & $\pm 0.10$ \\
$D^0\to K^+K^-\pi^+\pi^-$ background                               & $\pm 0.07$ \\
Selection of sideband                                              & $\pm 0.05$ \\
Variation of selection criteria                                    & $\pm 0.30$ \\
Fitting procedure                                                  & $\pm 0.10$ \\
Proper decay time range and binning                                & $\pm 0.07$ \\ 
Dalitz model                                                       & $\pm 0.01$\\
\hline
Total & $\pm 0.52$ \\
\hline 
\end{tabular}
\end{table*}

\section{Summary}
We present the first measurement of $y_{CP}$ using a $CP$-odd final state 
in $D^0$ decays. Our method has the advantage of not requiring precise knowledge 
of the decay-time resolution function, and avoids several biases that 
can arise due to detector effects.
The value of \ycp obtained is
\[
 y_{CP} = (+0.11\pm 0.61 ({\rm stat.})\pm 0.52 (\rm syst.))\%.
\]
This measurement of $y_{CP}$ using a $CP$-odd mode is consistent with previous measurements using $CP$-even final states~\cite{Staric,AubertMix}, and  with the world average value $y_{CP}=(+1.13\pm0.27)\%$ \cite{hfag}.

We thank the KEKB group for the excellent operation of the
accelerator, the KEK cryogenics group for the efficient
operation of the solenoid, and the KEK computer group and
the National Institute of Informatics for valuable computing
and SINET3 network support.  We acknowledge support from
the Ministry of Education, Culture, Sports, Science, and
Technology (MEXT) of Japan, the Japan Society for the 
Promotion of Science (JSPS), and the Tau-Lepton Physics 
Research Center of Nagoya University; 
the Australian Research Council and the Australian 
Department of Industry, Innovation, Science and Research;
the National Natural Science Foundation of China under
contract No.~10575109, 10775142, 10875115 and 10825524; 
the Department of Science and Technology of India; 
the BK21 program of the Ministry of Education of Korea, 
the CHEP src program and Basic Research program (grant 
No. R01-2008-000-10477-0) of the 
Korea Science and Engineering Foundation;
the Polish Ministry of Science and Higher Education;
the Ministry of Education and Science of the Russian
Federation and the Russian Federal Agency for Atomic Energy;
the Slovenian Research Agency;  the Swiss
National Science Foundation; the National Science Council
and the Ministry of Education of Taiwan; and the U.S.\
Department of Energy.
This work is supported by a Grant-in-Aid from MEXT for 
Science Research in a Priority Area ("New Development of 
Flavor Physics"), and from JSPS for Creative Scientific 
Research ("Evolution of Tau-lepton Physics").

\appendix*
\section{Integration of ${\cal A}_1{\cal A}^{\ast}_2$ over one Dalitz variable}
The amplitude ${\cal A}$ ($\overline{\cal A}$) for a $D^0$ ($\overline{D}{}^0$) decay to a three-body final state, $h^+h^-m^0$, depends on invariant masses of all possible pairs of final state particles: $s_0=M^2_{h^+h^-}$, $s_+=M^2_{h^+m^0}$ and $s_-=M^2_{h^-m^0}$. Only two of these three are independent, since energy and momentum conservation results in a constraint
\begin{equation}
 s_0+s_++s_-= m_{D^0}^2 + m_{h^+}^2 + m_{h^-}^2 + m^2_{m^0} \equiv  m^2.
\label{eq_con}
\end{equation}
In the limit of $CP$ symmetry the following relation holds:
\begin{eqnarray}
\overline{\cal A}(s_0,s_+) = {\cal A}(s_0,s_-) = {\cal A}(s_0, m^2 - s_+ - s_0),
\end{eqnarray}
and amplitudes ${\cal A}_{1,2}$ (defined in section~\ref{sec_method}) are then
\begin{equation}
 {\cal A}_1(s_0,s_+) =  \frac{1}{2}\left[{\cal A}(s_0,s_+)+{\cal A}(s_0, m^2 - s_+ - s_0)\right],
\end{equation}
\begin{equation}
 {\cal A}_2(s_0,s_+) =  \frac{1}{2}\left[{\cal A}(s_0,s_+)-{\cal A}(s_0, m^2 - s_+ - s_0)\right].
\end{equation}

We now show that
\begin{equation}
 \int_{s_+^{\rm min}(s_0)}^{s_+^{\rm max}(s_0)}{\cal A}_1(s_0,s_+){\cal A}_2^{\ast}(s_0,s_+)ds_+=0,
\label{eq_a1a2int}
\end{equation}
where $s_+^{\rm min}(s_0)$ and $s_+^{\rm max}(s_0)$ are lower and upper bounds of Dalitz variable $s_+$. For a given value of $s_0$, the range of $s_+$ is determined by its values when the 
momentum of $h^+$ is parallel or antiparallel to the momentum of $m^0$:
\begin{eqnarray}
 s_+^{\rm max}(s_0) & = & \left(E_{h^+}^{\ast}+E_{m^0}^{\ast}\right)^2\nonumber\\&&-\left(\sqrt{E_{h^+}^{\ast 2}-m_{h^+}^2}-\sqrt{E_{m^0}^{\ast 2}-m_{m^0}^2}\right)^2,~~~~~~\\
 s_+^{\rm min}(s_0) & = & \left(E_{h^+}^{\ast}+E_{m^0}^{\ast}\right)^2\nonumber\\&&-\left(\sqrt{E_{h^+}^{\ast 2}-m_{h^+}^2}+\sqrt{E_{m^0}^{\ast 2}-m_{m^0}^2}\right)^2,~~~~~~
\end{eqnarray}
where
\begin{eqnarray}
 E_{h^+}^{\ast} & = & \frac{s_0+m_{h^+}^2-m_{h^-}^2}{2\sqrt{s_0}},\\
 E_{m^0}^{\ast} & = & \frac{m_{D^0}^2-s_0+m_{m^0}^2}{2\sqrt{s_0}}
\end{eqnarray}
are the energies of $h^+$ and $m^0$ in the $h^+h^-$ rest frame.
The left-hand side of Eq.~(\ref{eq_a1a2int}) yields
\begin{eqnarray}
I&\equiv&\int_{s_+^{\rm min}(s_0)}^{s_+^{\rm max}(s_0)}{\cal A}_1(s_0,s_+){\cal A}_2^{\ast}(s_0,s_+)ds_+\nonumber\\ &=& \frac{1}{4}\left(I_a - I_b + I_c - I_d\right),
\label{eq_a1a2integrals}
\end{eqnarray}
where
\begin{equation}
 I_a = \int_{s_+^{\rm min}(s_0)}^{s_+^{\rm max}(s_0)} {\cal A}(s_0,s_+){\cal A}^{\ast}(s_0,s_+)ds_+,
\end{equation}
\begin{equation}
 I_b = \int_{s_+^{\rm min}(s_0)}^{s_+^{\rm max}(s_0)} {\cal A}(s_0,s_+){\cal A}^{\ast}(s_0, m^2 - s_+ - s_0)ds_+,
\end{equation}
\begin{equation}
 I_c = \int_{s_+^{\rm min}(s_0)}^{s_+^{\rm max}(s_0)} {\cal A}(s_0, m^2 - s_+ - s_0){\cal A}^{\ast}(s_0,s_+)ds_+,
\end{equation}
\begin{equation}
 I_d = \int_{s_+^{\rm min}(s_0)}^{s_+^{\rm max}(s_0)} {\cal A}(s_0, m^2 - s_+ - s_0){\cal A}^{\ast}(s_0, m^2 - s_+ - s_0)ds_+.
\end{equation}
In integrals $I_c$ and $I_d$ we perform a variable substitution $s_+\to s_-$ (Eq.~\ref{eq_con}):
\begin{subequations}
\label{eq_varsub}
\begin{equation}
  ds_+ = -ds_-,
\end{equation}
\begin{equation}
s_+^{\rm max}(s_0)\stackrel{m_{h^+}=m_{h^-}}{\longrightarrow}s_+^{\rm min}(s_0),
\end{equation}
\begin{equation}
s_+^{\rm min}(s_0)\stackrel{m_{h^+}=m_{h^-}}{\longrightarrow} s_+^{\rm max}(s_0),
\end{equation}
\end{subequations}
and obtain
\begin{equation}
 I_c = \int_{s_+^{\rm min}(s_0)}^{s_+^{\rm max}(s_0)} {\cal A}(s_0, s_-){\cal A}^{\ast}(s_0,m^2-s_--s_0)ds_- = I_b,
\end{equation}
\begin{equation}
I_d = \int_{s_+^{\rm min}(s_0)}^{s_+^{\rm max}(s_0)} {\cal A}(s_0, s_-){\cal A}^{\ast}(s_0,s_-)ds_- = I_a.
\end{equation}
The right-hand side of Eq.~(\ref{eq_a1a2integrals}) therefore yields zero.


\begin{thebibliography}{99}
\bibitem{CKM}
  N.~Cabibbo, Phys.\ Rev.\ Lett.\ {\bf 10}, 531 (1963);
  M.~Kobayashi, T.~Maskawa, Prog.\ Theor.\ Phys.\ {\bf 49}, 652 (1973).
\bibitem{GIM}
  S.L.~Glashow, J.~Illiopoulos, L.~Maiani, Phys.\ Rev.\ D{\bf 2}, 1285 (1970).
\bibitem{mix_th}
  I.I.~Bigi, N.~Uraltsev, Nucl.\ Phys.\ B {\bf 592}, 92 (2001); 
  A.F.~Falk, Y. Grossman, Z. Ligeti, A.A. Petrov, Phys.\ Rev.\ D{\bf 65}, 054034 (2002); 
  A.F.~Falk, Y. Grossman, Y. Nir, A.A. Petrov, Phys.\ Rev.\ D{\bf 69}, 114021 (2004).
\bibitem{Staric}
  M.~Staric {\it et al.}  [Belle Collaboration],
  Phys.\ Rev.\ Lett.\  {\bf 98}, 211803 (2007).
\bibitem{AubertMix}
  B.~Aubert {\it et al.}  [BABAR Collaboration],
  Phys.\ Rev.\ D{\bf 78}, 011105 (2008).
\bibitem{Aubert1}
  B.~Aubert {\it et al.}  [BABAR Collaboration],
  Phys.\ Rev.\ Lett.\  {\bf 98}, 211802 (2007).
\bibitem{CDF}
  T.~Aaltonen {\it et al.}  [CDF Collaboration],
  Phys.\ Rev.\ Lett.\  {\bf 100}, 121802 (2008).
\bibitem{kpipi0}
  B.Aubert {\it et al.}  [BABAR Collaboration],
  arXiv:0807.4544 [hep-ex], submitted to PRL.
\bibitem{hfag}
E.~Barberio {\it et al.}  [Heavy Flavor Averaging Group], arXiv:0808.1297 [hep-ex].
\bibitem{CC}
Throughout this paper, the inclusion of the charge-conjugate decay mode is 
implied unless stated otherwise.
\bibitem{MassConvention}
In the following, the nominal mass of particle $X$ is denoted as $m_X$, while the reconstructed invariant mass of system $Y$ is denoted as $M_Y$.
\bibitem{LiMing}
  L.M.~Zhang {\it et al.}  [Belle Collaboration],
  Phys.\ Rev.\ Lett.\  {\bf 99}, 131803 (2007).
\bibitem{AsnerCleo}
  D.~M.~Asner {\it et al.}  [CLEO Collaboration],
  Phys.\ Rev.\  D {\bf 72}, 012001 (2005).
\bibitem{BaBar}
  B.~Aubert {\it et al.}  [BABAR Collaboration],
  Phys.\ Rev.\  D {\bf 78}, 034023 (2008).
\bibitem{KEKB}
S.~Kurokawa and E.~Kikutani, Nucl. Instr. and. Meth. A {\bf 499}, 1 (2003),
and other papers included in this volume.
\bibitem{Belle}
A.~Abashian {\it et al.} (Belle Collaboration),
Nucl. Instr. and Meth. A {\bf 479}, 117 (2002).
 \bibitem{svd2}
 Z.~Natkaniec {\it et al.} (Belle SVD2 Group),
              Nucl. Instr. and Meth. A {\bf 560}, 1 (2006).
\bibitem{evtgen}
D.~J.~Lange, Nucl. Instr. and Meth. A {\bf 462}, 152 (2001).
\bibitem{GEANT}
R.~Brun {\it et. al.}, GEANT 3.21, CERN Report DD/EE/84-1 (1984).
\bibitem{BellePID}
  E.~Nakano,
  Nucl.\ Instr.\ and Meth.\  A {\bf 494}, 402 (2002).
\bibitem{BaBarII}
  B.~Aubert {\it et al.}  [BABAR Collaboration],
  Phys.\ Rev.\  D {\bf 72}, 052008 (2005).
\bibitem{model_diff}
In the Dalitz analysis of \zdecay decays~\cite{BaBarII}, the Dalitz model 
includes the $K^0_S a_0(980)^0$, $K^0_S \phi(1020)$, $K^0_S f_0(1370)$, 
$K^0_S f_0(980)$, and $K^- a_0(980)^+$ contributions. The fitted fractions 
of the latter two are consistent with zero, and the authors do not quote 
their amplitudes and phases, so these two contributions are not used in 
this paper. In the Dalitz analysis of Ref.~\cite{BaBar}, the Dalitz model 
includes the  $K^0_S a_0(980)^0$, $K^0_S \phi(1020)$, $K^0_S f_0(1370)$, 
$K^0_S f_2(1270)$, $K^0_S a_0(1450)^0$, $K^- a_0(980)^+$, $K^- a_0(1450)^+$, 
and $K^+ a_0(980)^-$ channels.
\end{thebibliography}
\end{document}